\documentclass{bioinfo}
\copyrightyear{2015}
\pubyear{2015}

\usepackage{amsmath}
\usepackage{natbib}

\begin{document}
\firstpage{1}

\title[FermiKit: assembly based variant calling]{FermiKit: assembly-based variant calling for Illumina resequencing data}

\author[Li]{Heng Li}

\address{Broad Institute, 75 Ames Street, Cambridge, MA 02142, USA}

\history{Received on XXXXX; revised on XXXXX; accepted on XXXXX}
\editor{Associate Editor: XXXXXXX}
\maketitle

\begin{abstract}
\section{Summary:}
FermiKit is a variant calling pipeline for Illumina data. It {\it de novo}
assembles short reads and then maps the assembly against a reference genome to
call SNPs, short insertions/deletions (INDELs) and structural variations (SVs).
FermiKit takes about one day to assemble 30-fold human whole-genome data on a
modern 16-core server with 85GB RAM at the peak, and calls variants in half an hour to
an accuracy comparable to the current practice. FermiKit assembly is a reduced
representation of raw data while retaining most of the original information.

\section{Availability and implementation:} https://github.com/lh3/fermikit

\section{Contact:} hengli@broadinstitute.org
\end{abstract}

\section{Introduction}
Deep resequencing of a human sample typically results in a BAM file of
60--100GB in size. Storing, distributing and processing many such huge files
is becoming a burden for sequencing facilities and research labs. While better
compression helps to alleviate this issue, it adds processing time and can
barely halve the size, which does not keep up with the rapidly increasing
sequencing throughput. Illumina and GATK use gVCF~\citep{Raczy:2013aa} as a
reduced representation of raw data. However, gVCF is reference dependent and it
is nontrivial to encode both large and small variants consistently. We still
need to go back to raw data for long events and when upgrading the reference
genome. Another idea from the past practice is to assemble sequence reads into
contigs that ideally retains all information in the raw data, but whether this
approach is practical to Illumina human resequencing remains to be confirmed.

\begin{methods}
\section{Methods}
The FermiKit pipeline uses BFC~\citep{Li:2015aa-tmp} for error correction,
ropeBWT2~\citep{Li:2014ab} for BWT construction, an improved version of
fermi~\citep{Li:2012fk} for {\it de novo} assembly, BWA-MEM~\citep{Li:2013aa}
for mapping and HTSBox (http://bit.ly/HTSBox) for variant calling. The
assembler retains the number of supporting reads at each position on the
contigs. The caller simply parses edits in the `pileup' output for small
variant calling from one or multiple BAMs, and extracts alignment break points
for SV calling. It sets thresholds on mapping quality and the number of
supporting reads without using sophisticated statistical models. FermiKit
does not use paired-end information for the time being, but this does not have
a great impact on its power empirically. With longer upcoming Illumina reads,
it will be actually preferred to merge overlapping ends and treat them as
single-end reads.

\end{methods}

\begin{table}[t]
\processtable{GIAB evaluation on SNP/INDEL accuracy}
{\footnotesize
\begin{tabular}{lp{1.7cm}rrrr}
\toprule
Sample & Caller     & SNP-FN & SNP-FP & InDel-FN & InDel-FP \\
\midrule
PG- & FermiKit      & 45,700 & 824    & 2,324    & 472 \\
	& FreeBayes     & 21,601 & 440    & 3,856    & 616 \\
    & HC+hardFilter & 27,010 & 144    & 943      & 370 \\
	& HC+VQSR       & 128,604& 1,955  & 1,423    & 366 \\
S7- & FermiKit      & 65,217 & 531    & 2,340    & 549 \\
	& FreeBayes     & 50,866 & 675    & 2,895    & 670 \\
    & HC+hardFilter & 66,847 & 228    & 1,543    & 457 \\
	& HC+VQSR       & 103,979& 1,508  & 1,396    & 605 \\
S11-& FermiKit      & 91,468 & 541    & 2,973    & 554 \\
	& FreeBayes     & 52,120 & 904    & 3,200    & 672 \\
	& HC+hardFilter & 65,223 & 407    & 1,502    & 472 \\
	& HC+VQSR       & 111,504& 1,694  & 1,175    & 765 \\
S2- & FermiKit      & 63,445 & 448    & 2,244    & 568 \\
S12-& FermiKit      & 74,940 & 501    & 2,562    & 553 \\
S1+ & FermiKit      & 67,816 & 455    & 4,051    & 516 \\
	& FreeBayes     & 63,147 & 902    & 4,626    & 660 \\
    & HC+hardFilter & 71,174 & 531    & 2,376    & 591 \\
	& HC+VQSR       & 108,101& 8,852  & 2,377    & 1,827 \\
S4+ & FermiKit      & 71,262 & 452    & 4,197    & 536 \\
	& FreeBayes     & 65,476 & 1,057  & 4,782    & 661 \\
	& HC+hardFilter & 75,040 & 672    & 2,477    & 653 \\
	& HC+VQSR       & 103,595& 10,492 & 2,401    & 1,622 \\
\botrule
\end{tabular}}{PCR-free Platinum Genome NA12878 (PG-; AC:ERR194147), four Illumina X10 lanes
of PCR-free NA12878 (S7-, S11-, S2- and S12- under BaseSpace project ID
18475457) and two X10 lanes of PCR-amplified NA12878 (S1+ and S4+ under project
ID 8998991) were acquired and called with FermiKit-0.9, FreeBayes-0.9.20
(option: `--experimental-gls --min-repeat-entropy 1') and HC-3.3 (option:
`-stand\_emit\_conf 10 -stand\_call\_conf 30'). For FreeBayes and HC, BWA-MEM
was used for mapping with PCR duplicates marked by
Samblaster~\citep{Faust:2014aa}.  Short variant calls were hard filtered with
hapdip (http://bit.ly/HapDip). GATK-VQSR was also applied to HC calls. The
filtered calls were compared to GIAB-v2.18 excluding poly-A regions longer than
6bp plus 10bp flanking. A true variant is counted as an FN if there are no
called variants within 10bp around the truth, and a called variant is counted
as an FP if it falls in GIAB trusted regions and there are no true variants
within 10bp around the called variant.}
\end{table}

\section{Results}
We have run FermiKit on multiple NA12878 whole-genome data sets along with
GATK-HaplotypeCaller (HC in brief) and FreeBayes~\citep{Garrison:2012aa}.
We used Genome-In-A-Bottle (GIAB; \citealt{Zook:2014ab}) as truth data
to evaluate the accuracy (Table~1). Recent Illumia data have excessive
systematic errors around poly-A which HC does not handle well. It called over
4000 false INDELs from sample S1+ and S4+ with the vast majority around poly-A.
We excluded these regions to avoid one simple error source greatly affecting
the metrics. After this treatment, variant callers are broadly comparable when
the same set of hard filters are applied. VQSR as is advised in GATK Best
Practice does not work well with single-sample calling.

GIAB was generated from multiple NA12878 call sets. It is potentially biased
against new callers and biased towards easier regions that can be called by the
existing callers. For example, the GATK call set available from the Platinum
Genome website has 13,278 FN SNPs and 46 FPs out of 2.03Gb confident regions
(i.e. one SNP error per 44Mbp), which is overly good and is worrying.

\begin{table}[t]
\processtable{Evaluation on SNP/INDEL accuracy with CHM1-NA12878 pair}
{\footnotesize
\begin{tabular}{llrrrr}
\toprule
Caller   & Filter     & SNP-TP    & SNP-FP & InDel-TP & InDel-FP \\
\midrule
FermiKit & hard-polyA & 1,937,469 & 22,743 & 230,955 & 14,602 \\
         & uniMask    & 1,802,820 & 9,507  & 127,304 & 1,126 \\
FreeBayes& hard-polyA & 2,026,883 & 59,422 & 190,587 & 30,909 \\
         & uniMask    & 1,842,634 & 15,252 & 117,764 & 6,329 \\
HC       & hard-polyA & 2,003,655 & 32,030 & 267,870 & 15,541 \\
         & uniMask    & 1,824,658 & 14,912 & 133,458 & 2,046 \\
\botrule
\end{tabular}}{SNP/INDELs were called from the CHM1 (AC:SRR642636 through SRR642641) and NA12878-PG- BWA-MEM
alignments used by~\citet{Li:2014aa}. On the assumption that CHM1 is haploid,
(heterozygous) FP equals the number of CHM1 heterozygotes and (heterozygous) TP
equals the number of NA12878 heterozygotes minus the number of CHM1
heterozygotes. Two sets of filters were applied for filtering. `Hard-polyA' is
the same as the filter used in Table~1.  `UniMask' filters out genomic regions
that tend to be repetitive, low-complexity or susceptible to copy number
changes or systematic artefacts (http://bit.ly/unimask). This filter is
sample independent.}
\end{table}

\begin{table}[t]
\processtable{Performance on calling long deletions over 100bp}
{\footnotesize
\begin{tabular}{llcccc}
\toprule
Sample& Caller  & {1000g pilot} & {Ensemble} & {LUMPY} & {Merged}\\
\midrule
S7- & FermiKit  & 0.43 / 0.23 & 0.50 / 0.15 & 0.32 / 0.23 & 0.58 / 0.09 \\
S1+ & FermiKit  & 0.43 / 0.22 & 0.51 / 0.15 & 0.33 / 0.23 & 0.58 / 0.10 \\
PG- & FermiKit  & 0.43 / 0.20 & 0.52 / 0.14 & 0.34 / 0.22 & 0.59 / 0.09\\
    & DELLY     & 0.47 / 0.34 & 0.50 / 0.22 & 0.31 / 0.28 & 0.58 / 0.16\\
    & LUMPY     & 0.72 / 0.34 & 0.76 / 0.29 & 0.68 / 0.37 & 0.79 / 0.20\\
\botrule
\end{tabular}}{FermiKit was used to call 100bp or longer deletions from the
PG-, S7- and S1+ datasets. DELLY~\citep{Rausch:2012aa} and LUMPY PG- calls were
acquired from http://bit.ly/bcbsval (B. Chapman, personal communication). For
all call sets, overlapping events were merged and deletions longer than 100kbp
were discarded.  The two numbers in a cell at row $R$ and column $C$ give the
false negative rate and false positive rate of call set $R$, assuming truth set
$C$ is correct and complete.  In the table, truth set `1000g pilot' consists of
deletions by \citet{Mills:2011aa} and further validated
by~\citet{Layer:2014aa}; `Ensemble' contains validated calls by multiple
callers; `LUMPY' consists of validated LUMPY-only deletions; `Merged' is the
union of all the three truth sets above.}
\end{table}

We turned to the CHM1-NA12878 dataset~\citep{Li:2014aa} for an unbiased
evaluation (Table~2). In this evaluation, FermiKit produces calls of higher
specificity at the cost of sensitivity. This is probably because FermiKit is
less powerful in repetitive or duplicated regions or regions affected by
systematic artefacts. Nonetheless, in well-behaved regions that are outside
`uniMask', the loss of sensitivity is minor. The gain in precision is
significant if we consider that there may be 5--20k real heterozygous SNPs in
CHM1~\citep{Li:2014aa}, which should not be counted as FPs.

FermiKit performs well in calling long deletions (Table~3). While
it does not use read pairs, it achieves comparable sensitivity and higher
specificity in comparison to the state of art. FermiKit also called $\sim$900
novel sequence insertions and identified multiple kb-long contigs having 
poor alignments to GRCh37 but nearly perfect alignment to a PacBio assembly of
CHM1 (J. Chin, personal communication). We also mapped the CHM1 unitigs to
this assembly and called 71 long deletions, 35 insertions and 262 other events.
As PacBio assemblies are generally of higher quality, these numbers give a
rough estimate on the number false positives for haploid data.

\section{Conclusions}

A FermiKit assembly is about 3GB compressed. After assembly, single-sample
variants can be obtained in half an hour to high accuracy through mapping
against a reference genome. Jointly calling 261 aligned SGDP samples only took
$\sim$40 CPU hours.  FermiKit is a viable option for aggressive data
compression, greatly reducing the efforts and expense on data storage,
distribution and re-analyses at an acceptable cost of information loss.

\section*{Acknowledgement}
\paragraph{Funding\textcolon} NHGRI U54HG003037; NIH GM100233

\bibliography{fermikit}

\begin{thebibliography}{}

\bibitem[Faust and Hall, 2014]{Faust:2014aa}
Faust, G.~G. and Hall, I.~M. (2014).
\newblock Samblaster: fast duplicate marking and structural variant read
  extraction.
\newblock {\em Bioinformatics}, 30:2503--5.

\bibitem[Garrison and Marth, 2012]{Garrison:2012aa}
Garrison, E. and Marth, G. (2012).
\newblock Haplotype-based variant detection from short-read sequencing.
\newblock {\em arXiv:1207.3907}.

\bibitem[Layer et~al., 2014]{Layer:2014aa}
Layer, R.~M. et~al. (2014).
\newblock {LUMPY}: a probabilistic framework for structural variant discovery.
\newblock {\em Genome Biol}, 15:R84.

\bibitem[Li, 2012]{Li:2012fk}
Li, H. (2012).
\newblock Exploring single-sample {SNP} and {INDEL} calling with whole-genome
  de novo assembly.
\newblock {\em Bioinformatics}, 28:1838--44.

\bibitem[Li, 2013]{Li:2013aa}
Li, H. (2013).
\newblock Aligning sequence reads, clone sequences and assembly contigs with
  {BWA-MEM}.
\newblock {\em arXiv:1303.3997}.

\bibitem[Li, 2014a]{Li:2014ab}
Li, H. (2014a).
\newblock Fast construction of {FM-index} for long sequence reads.
\newblock {\em Bioinformatics}, 30:3274--5.

\bibitem[Li, 2014b]{Li:2014aa}
Li, H. (2014b).
\newblock Toward better understanding of artifacts in variant calling from
  high-coverage samples.
\newblock {\em Bioinformatics}, 30:2843--2851.

\bibitem[Li, 2015]{Li:2015aa-tmp}
Li, H. (2015).
\newblock {BFC}: correcting illumina sequencing errors.
\newblock {\em arXiv:1502.03744}.

\bibitem[Mills et~al., 2011]{Mills:2011aa}
Mills, R.~E. et~al. (2011).
\newblock Mapping copy number variation by population-scale genome sequencing.
\newblock {\em Nature}, 470:59--65.

\bibitem[Raczy et~al., 2013]{Raczy:2013aa}
Raczy, C. et~al. (2013).
\newblock Isaac: ultra-fast whole-genome secondary analysis on illumina
  sequencing platforms.
\newblock {\em Bioinformatics}, 29:2041--3.

\bibitem[Rausch et~al., 2012]{Rausch:2012aa}
Rausch, T. et~al. (2012).
\newblock {DELLY}: structural variant discovery by integrated paired-end and
  split-read analysis.
\newblock {\em Bioinformatics}, 28(18):i333--i339.

\bibitem[Zook et~al., 2014]{Zook:2014ab}
Zook, J.~M. et~al. (2014).
\newblock Integrating human sequence data sets provides a resource of benchmark
  snp and indel genotype calls.
\newblock {\em Nat Biotechnol}, 32:246--51.

\end{thebibliography}
\end{document}